# BCNet: A Deep Convolutional Neural Network for Breast Cancer Grading

Pouya Hallaj Zavareh[1], Atefeh Safayari[2], Hamidreza Bolhasani[3]

*Abstract*— **Breast cancer has become one of the most prevalent cancers by which people all over the world are affected and is posed serious threats to human beings, in a particular woman. In order to provide effective treatment or prevention of this cancer, disease diagnosis in the early stages would be of high importance. There have been various methods to detect this disorder in which using images have to play a dominant role. Deep learning has been recently adopted widely in different areas of science, especially medicine. In breast cancer detection problems, some diverse deep learning techniques have been developed on different datasets and resulted in good accuracy. In this article, we aimed to present a deep neural network model to classify histopathological images from the Databiox image dataset as the first application on this image database. Our proposed model named BCNet has taken advantage of the transfer learning approach in which VGG16 is selected from available pertained models as a feature extractor. Furthermore, to address the problem of insufficient data, we employed the data augmentation technique to expand the input dataset. All implementations in this research, ranging from pre-processing actions to depicting the diagram of the model architecture, have been carried out using tf.keras API. As a consequence of the proposed model execution, the significant validation accuracy of 88% and evaluation accuracy of 72% obtained.**

## 1. Introduction

Breast cancer has been recently announced as one of the most common cancers amongst people, especially women [1]. According to statistics, in 2020, approximately 1.8 million new cases were diagnosed with all types of cancer of which more than a half belonged to women among who more than 30% suffered from breast cancer [2]. This disease has become one of the important concerns to women because it is second only to lung cancer in cancer death among women [3].

Early detection considers being of primary importance in the effective and efficient treatment of all disorders. This cancer can be prevented or administered if it is diagnosed in time [4]. To achieve this matter, various researches have been conducted on this subject [5] which aimed to fulfill the purpose of early diagnosis using different methods such as mammography [6], Microwave imaging (MI) [7], and Magnetic resonance imaging (MRI) [8]. Deep learning has recently been utilized as a beneficial tool to identify affected cases [9, 10].

[1] Department of Computer Engineering, Islamic Azad University, Isfahan, Iran.
[2] Department of Computer Science, Kharazmi University of Tehran, Tehran, Iran.
[3] Department of Computer Engineering, Islamic Azad University Science and Research Branch, Tehran, Iran.
+ Corresponding Author: hamidreza.bolhasani@srbiau.ac.ir

Deep learning is a hierarchical structure network that is comprised of various algorithms in which hidden neurons are considered the key parts of architecture [11]. Due to this construction, they are capable of deriving complex concepts from simple statements [12]. These extracted features with the assistance of multilayer structure pursue the goal of pattern recognition from input data in an end-to-end way to resolve the problem [13]. Fig.1 exhibits a sample of a DNN.

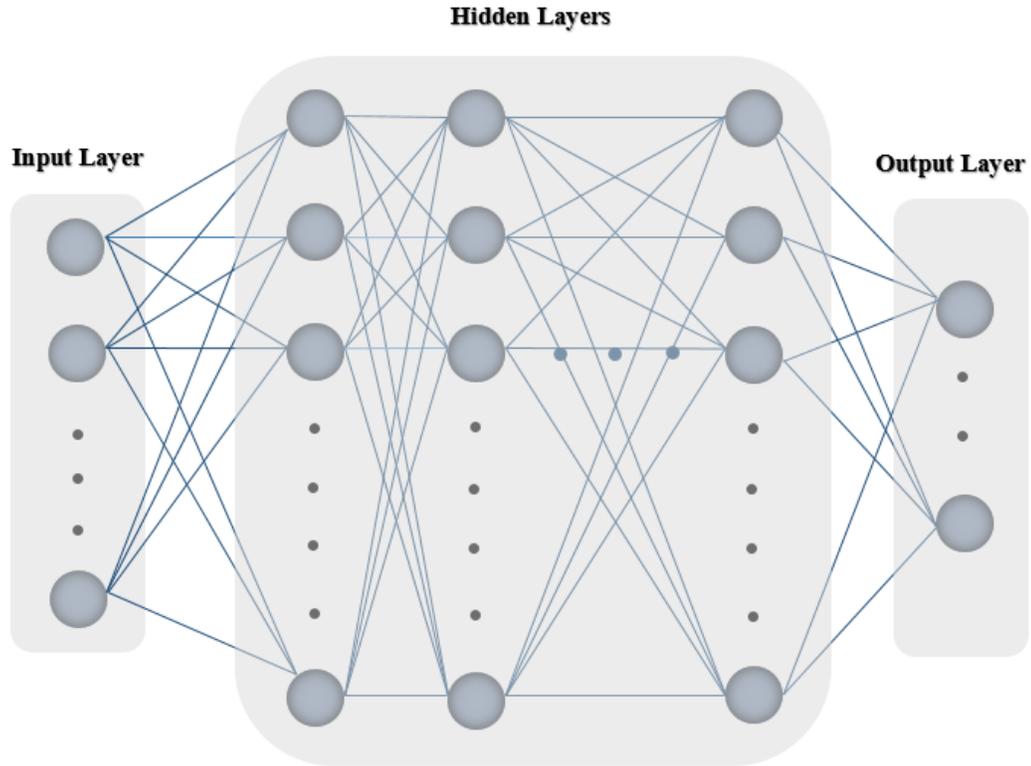

Fig.1 Deep learning network structure

Although deep learning has been applied in a diverse range of applications of different areas such as the automotive Industry [14-15], and IoT [16], its usage in medicine has been newly received more attention [17, 19]. The layer-based architecture of deep learning with the ability to automate learning has provided the medical applications with the state-of-the-art solution and result to detect or predict disease [20]. Adopting images to diagnose human cancer like breast cancer, which is known as cancer imaging, has resulted in considerable success in cancer detection [21].

With respect to breast cancer, there have been several applications of deep learning in the classification of breast cancer images using deep learning. Breast cancer histological images are widely employed to detect this type of cancer. In this study, we aimed to grade some real histological images with the assistance of deep neural networks. After some investigation and examination, a transfer learning approach was applied in this research in which the VGG16 model as a very deep pre-trained neural network was utilized to extract desired features. All processes carried out using utilities of tf.keras API. To the best of our knowledge, there was no application on the adopted image dataset in our exploration and our proposed model is the pioneer in using this dataset to grade and classify.
The important contributions of this study are as follow:

- Presenting a deep learning model on employed image dataset whose information is provided in section 3.1

- Providing schematic diagrams of used methods in each step such as pre-processing, and model architecture

In the organization of this research, related works are discussed in **section 2**. In **section 3** the proposed model, training process, and result is presented. The conclusion is provided in **section 4**. **Section 5**, dedicates to explain the contributions of authors.

2. Relate works

Diverse range of researches has been conducted on breast cancer histopathological images classification and grading. In this section we intend to explain some related studies to our objective. Some important features of them are illustrated in table 1.

Boumaraf et al. [22] classified BreaKHis histopathological images [23] in two different ways that were binary and eight-class groups. Binary classification aimed to categories the specified medical images, which had the same level of magnification, into two Magnification Dependent (MD) and Magnification Independent (MI) binary classes. Classifying images into eight groups was also the purpose of the second classification. By adopting the global contrast normalization (GCN) technique, the image dataset normalized, and then a three-fold data augmentation was employed to augment the training data. These preparing tasks addressed overfitting reduction and performance improvement. Transfer learning followed by a block-wise fine-tuning strategy utilized for the training process. This action was performed with the assistance of the ResNet-18 network from which desired features were extracted and used in the final architecture. This pre-trained module in which there were two residual blocks. This study took advantage of obtaining a considerable accuracy rather than previous works in both automatic classification types.

Kumar et al. [24] conducted classifications on two different image datasets that one of them was BreakHis, a histopathology image dataset of human breast cancer. Adopting Keras libraries, the specified firstly normalized. The final pre-processing performed through a two-part step contained resizing and augmentation. To train data, a combined framework of deep learning and machine learning methods presented. The proposed model was comprised of two general sections, feature extraction and classification. VGGNet-16 was employed as a feature generator and aimed to take appropriate features into the classification layer as a strategy called transfer learning. To do this, after removing the last fully connected layer and adding a global average pooling (GAP) at the end of each extraction block, a feature vector of produced features built that fed into the classifier. Various algorithms such as SVM, Random Forest (RF) were exploited to carry out the categorization between which VGGNet-16 with RF using images that were magnified 200 times gained the best accuracy.

Alom et al. [25] classified histopathological images from two BreakHis and Breast Cancer (BC) datasets. Both binary and multi-class classifications were addressed in this study. Considering the multi-class approach, two four and eight classes were deemed. To overcome overfitting due to small data, first of all, input images were downsampled using the augmentation technique. Output images were then evaluated using the center patch method. The designed model which is called Inception Recurrent Residual Convolutional Neural Network (IRRCNN) was a combination of the Inception Network (Inception-v4), the Residual Network (ResNet), and the Recurrent Convolutional Neural Network (RCNN). This proposed DCNN model. Model performance was evaluated using four analyses, namely, image-level, patient-level, image-based, and patch-based. Images classified in this study are based on various criteria such as magnification factor, resized sample inputs, augmented patches and samples, and patch-based. The incredible result of 100% was obtained by adopting random patches and Winner Take All (WTA) approaches for image-based recognition in order to evaluate the performance of the designed model.

Employment of different methods and Criteria and providing classification for each of them in pair considered the strong point of this research.

Golatkar et al. [26] aimed to classify breast cancer images that were stained with H&E into four normal tissue, benign lesion, *in situ* carcinoma, and invasive carcinoma classes. The used histology microscopy image dataset was generated for the Breast Cancer Histology Challenge (BACH) [27] To avoid overfitting due to lack of data, they proposed a technique called Nuclei based patch extraction to take patches based on nuclear density in which 299 by 299 pixels patches with a 50% overlap were extracted by the division of images. As a consequence, just patches with a high density of nuclear were selected to prepare and those that were mostly covered by stoma excluded. The output patches of this step are then followed by some preprocessing and data augmentation actions. They fulfilled the goal of training and classification utilizing transfer learning and fine-tuning strategies. They proposed a customized deep convolutional neural network using Inception-v3 as the base of the architecture. The last fully-connected layer was replaced by three layers which were a global average pooling layer, a fully connected layer with 1,024 neurons, and a softmax classifier with 4 neurons. RMSProp, SGD, and momentum were three used optimizers by this study during three different stages of the training process. To evaluate the model a twofold validation was applied. This research took advantage of considering only relevant regions of images to train and test the model. Lack of data and *region-of-interest annotations* were deemed two major problems.

The breast cancer histology images classification into four normal, benign, *in situ*, and invasive categories in [28] carried out using images from [29] which were provided in the ICIAR 2018 grand challenge on breast cancer histology images. To prepare data, after normalization as first step, images augmented and cropped. Then, using 3-norm pooling method, a single deep neural network descriptor created from a combination of twenty descriptors by which several datasets based on crop sizes, encoders, and number of scales built. This study practiced transfer learning strategy to classify images without applying fine-tunning. Three deep pre-trained convolutional neural networks named ResNet-50, InceptionV3 and VGG-16 customized to extract desired features and train the dataset. Instead of the last fully connected layer, in both ResNet-50, InceptionV3 networks, a one-dimensional feature vector was put. In VGG-16 network, a GlobalAveragePooling layer added into each of two to five blocks and then these added layers gather into a feature vector. LightGBM employed as optimizer and model evaluation done by adopting 10-fold cross validation method. This research also suffer from insufficient data.

A binary classification into two classes benign (B) and malignant (M) carried out by [30] on the BreakHis histology image dataset. This research provided a remarkable comparison between using pre-trained networks and training from scratch. The used data pre-processed using Keras utilities. In the full-training approach, they randomly initialized weights, while, in the second technique, the pre-trained networks were utilized as the core of the proposed model. With respect to using transfer learning to achieve the goal of training, three pre-trained deep convolutional neural network models namely VGG16, VGG19, and ResNet50 adopted as feature extractors by which the generated features were used in the classification step. Logistic regression (LR) aimed to determine the final decision as a classifier whose performance was measured by a receiver operating characteristics (ROC) analysis. Furthermore, an area under the curve (AUC) and average precision score (APS) were respectively responsible for validating the result and performance evaluation. In the end, it was observed that using pre-trained modules, VGG16 and VGG-19 in particular, in the context of transfer learning employment as the base of the designed model will be led to a more accurate value.

Table. 1. Main information of studied researches on classification histopathological images of breast cancer

| Reference | Publication year | Method | Utilized Dataset |
|---|---|---|---|
| Boumaraf et al. [22] | 2020 | Transfer learning and fine-tuning with ResNet-18 | BreaKHis [23] |
| Kumar et al. [24] | 2019 | - Transfer learning with VGGNet-16 as a feature generator<br>- Adopting SVM, Random Forest (RF) for classification | BreaKHis [23] |
| Alom et al. [25] | 2019 | Inception Recurrent Residual Convolutional Neural Network (IRRCNN) | BreakHis [23] and Breast Cancer (BC) |
| Golatkar et al. [26] | 2018 | Transfer learning and fine-tuning with Inception-v3 | the Breast Cancer Histology Challenge (BACH) [27] |
| Rakhlin et al. [28] | 2018 | Transfer learning without fine-tuning using Inception-v3, ResNet-50, VGG16 | ICIAR challenge [29] |
| Mehra et al. [30] | 2018 | - Transfer learning with VGG16, VGG19, and ResNet50 as a feature generator<br>- Adopting Logistic regression (LR) for classification | BreaKHis [23] |
| Our research | 2021 | Transfer learning without fine-tuning using VGG16 | Databiox [31] |

### 3. Methods

In this section, the detail of our adopted method and the proposed model is being described. Firstly we delineate the employed image dataset in this study. This follows by some explanation about pre-processing step. Then, the architecture of our model expands in the following part, and some information about training and the result of the presented model is provided in the two next sections.

**3.1 Dataset**

The utilized dataset in this paper has been provided by [31]. This histopathological microscopy image dataset named Databiox is comprised of 922 stained images with Hematoxylin and eosin (H&E) which has been captured from breast tissues of 124 patients who suffered from IDC. These images are prepared in RGB and JPEG format with the resolution of 2100 by 1574 and 1276 by 956 pixels and 72 DPI. All specimens have been presented with four 4x, 10x, 20x, and 40x microscopy magnification levels. A modified histologic grading method called Bloom Richardson has been adopted to carry out the process of specimen labeling by which each sample is labeled with a class among three defined grade I, grade II, and grade III categories. Grade I means well-differentiated, grade II denotes moderately differentiated, and poorly differentiated is indicated by grade III. An example of each grad is given in fig.1. This introduced dataset offers the advantage of the same numbers of specimens for each proposed classification of IDC grading.

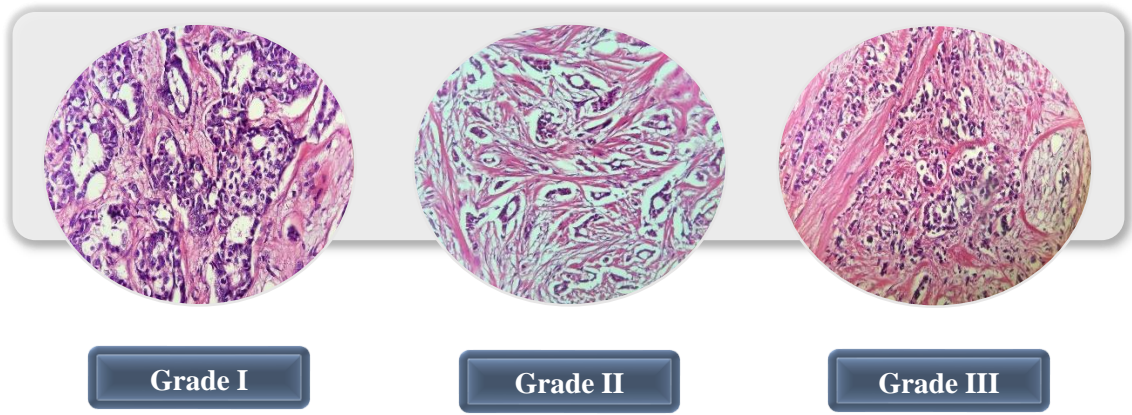

Grade I    Grade II    Grade III

Fig.2 Samples for each grade

## 3.2 Approach overview

Lack of a dataset of sufficient size considers a major problem by which serious difficulties are posed to the training process of a deep learning model [32]. Overfitting [33] and being unable to train a convolutional neural network from scratch [34] would be two direct consequences of inadequate numbers of images. Transfer learning, which includes leveraging learned features from a problem on a new, similar one, is a technique that is aimed to compensate for the insufficient training data by the employment of a pre-trained model [35, 36]. It can be followed by fine-tuning which is defined as a transfer learning approach that is prevalent in deep learning applications [37]. Although fine-tuning can provide the model with significant improvement, this optional step can also lead to overfitting [38]. With respect to this fact and also considering our experiments, this paper has only taken advantage of transfer learning and pre-trained, very deep models. Considerable numbers of computer vision tasks and problems have obtained the state-of-the-art results with the assistance of utilizing some pre-trained deep convolutional neural networks such as Google Net [39], Xception [40], ResNet50 [41], InceptionV3 [42] VGG16, InceptionV4 [43]. Among the available models which some have been mentioned, VGG16 [44] as the winner of the ImageNet Large-Scale Visual Recognition Challenge (ILSVRC) [45] is a very deep convolutional neural network that has trained on the large image dataset named ImageNet [46]. As VGG16 outperformed other models in our specific research, our final model, which is named BCNet, has been constructed based on this ConvNets architecture and has been applied to the pre-processed histopathological microscopy image dataset to be analyzed and encoded. Lastly, this procedure ended up with practicing a gradient-based optimization algorithm [47] called adaptive moment estimation (Adam) [48] to optimize the performance and loss value. We employed a deep learning framework and library called TensorFlow [49, 50] and utilized tf.keras API which is integrated form of Keras in it [51]. Fig.3 offers a broad overview of adopted approach in this study.

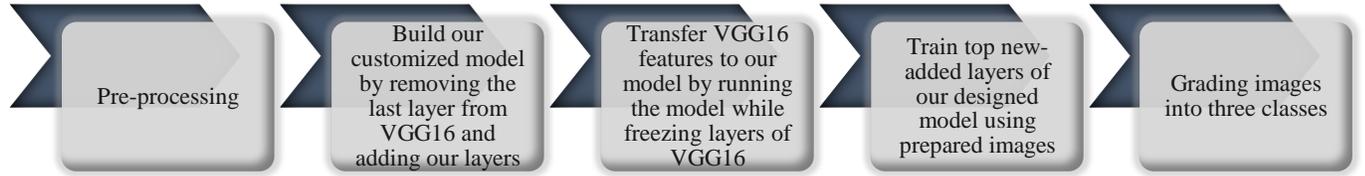

Fig.3 General overview of applied method

### 3.3 Data pre-processing and augmentation

In order to provide the training process with suitable data to feed, we made use of dataset preprocessing utilities that are available in tf.keras API which is mentioned in section 3.2. Prior to the commencement of data preprocessing, we randomly-manually selected 21 images to equip the evaluation step with required data and the rest of the images, later, divided into two training and validation groups using mentioned utilities of Keras. Data preparation was carried out through several steps. To begin with, we aimed to address the problem of overfitting, which stems from limited data [52], by the employment of a regularization technique termed data augmentation to increase the size of the input image dataset by generating more images with slight difference from the original one [53, 54]. We utilized ImageDataGenerator class which is built to fulfill the purpose of data augmentation in tf.keras API **[55]**. The result of applying various modifications on images finally led us to only selecting three arguments to alter them which were preprocessing_function, zoom_range, and validation_split parameters with values preprocess_input function, 0.2, and 0.25 respectively. Preprocess_input function is the specific input processing of VGG16 by which images are transformed from RGB format to BGR format, followed by the zero-center process of each color channel without scaling. Then these changes were applied to all images of the training dataset using the flow_from_directoy method in the form of iterations include 52 images for each repetition **[56]**. Furthermore, in each specified iteration, images were resized to 224 by 224 pixels. This procedure followed by carrying out conversion to RGB format and data shuffling was enabled by setting the shuffle parameter to true. Moreover, the type of label of returned arrays selected categorical to produce 2D one-hot encoded labels. The detail of two used utilities is provided in table 2. In addition, fig.4 displays a comprehensive overview of pre-processing operations.

Table 2. Detail explanation of used tf.keras utilizes to prepare image dataset

| tf.keras Dataset Preprocessing Utility Name | Parameter Name | Value |
|---|---|---|
| ImageDataGenerator | Preprocessing_function | preprocess_input |
|  | Zoom_range | 0.2 |
|  | Validation_split | 0.25 |
| Flow_from_directoy | Target_size | (224,224) |
|  | Color_mode | rgb |
|  | Batch_size | 52 |
|  | Class_mode | categorical |

|  | Shuffle | TRUE |
| --- | --- | --- |

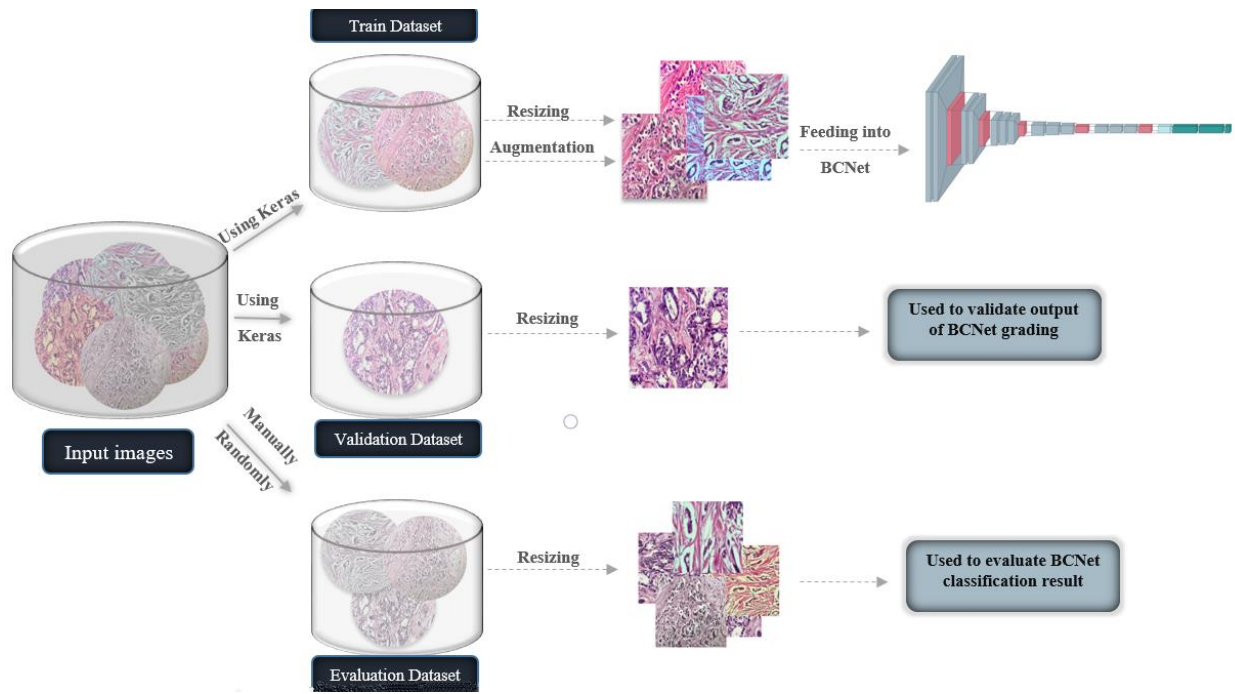

Fig.4 Detailed illustration of pre-processing step

### 3.4 Feature Extraction

With respect to our proposed model which is constructed based on transfer learning technique, and also considering reasons of the pre-trained model selection for this research, VGG16 network adopted to extract required features [57]. To fulfill the goal of feature extraction, firstly, an instance of VGG16 as a base model was created in which the three fully connected layers at the top of the network which was contained 1000 nodes was removed by setting the include-top parameter to false. The pre-trained weights then were loaded into it. To create a new model, we aimed to add some layers on the top of the base model, so this model froze by assigning false value to its trainable parameter. In this study, the mentioned output part was comprised of one GlobalAveragePooling2D layer with 512 nodes, two dense layers with 1024 nodes and ReLU activation function, and another dense layer as classification layer with 3 nodes and Softmax function. At the end, by running the designed model, the feature extraction process took place. Generally, the proposed deep convolutional neural network is consisted of 6 blocks between which the last one belongs to added customized layers. Fig 5 exhibits the structure of the presented model. Visualkeras utility from Keras was employed to depict the represented picture [58] in figure 5.

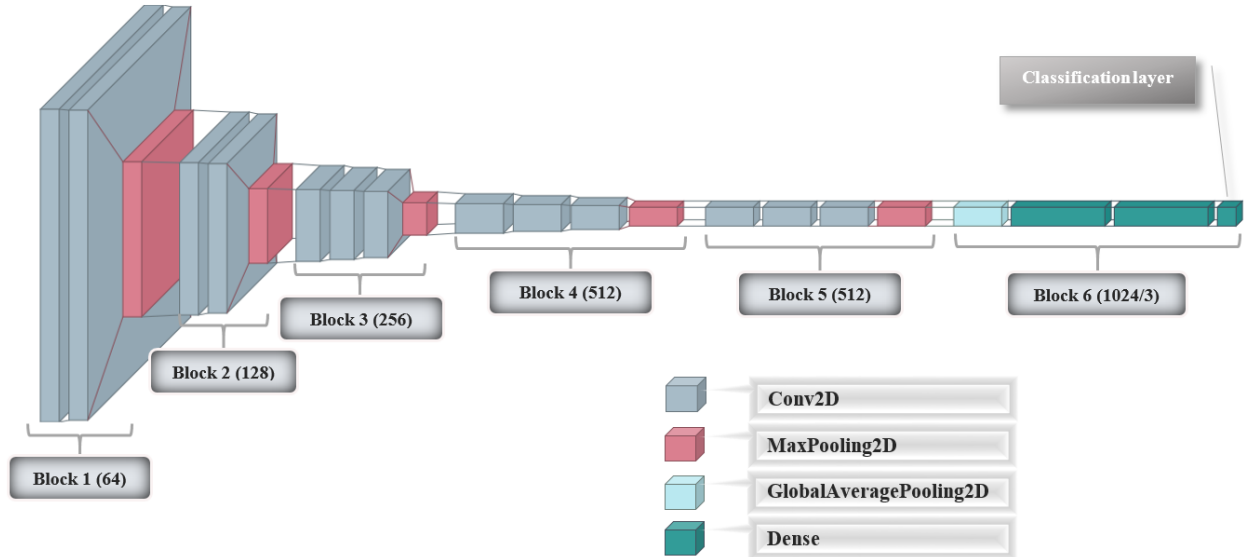

Fig.5 The architecture of the proposed deep learning model

### 3.5 Training

The general structure of our designed model called BCNet is displayed in fig.5 and its design and creation steps are expounded in section 4.4. Respecting data splitting, as it pointed out in section 3.3, prepared data was partitioned into two training and validation portions by which 75% of data dedicated to the training process. Begin with, by using compile() function, the mentioned CNN architecture created as a python object in which Adam selected as the optimizer, loss function decided to be categorical_crossentropy, and metrics parameter set to accuracy. This model configuration step, then followed by the employment of fit function to train the top layer of our model by applying it on the pre-processed training dataset considering the prepared validation data. This was carried out throughout 50 epochs. Furthermore, in each stage of the training process, a custom callback called to stop training if the obtained accuracy of validation action exceeds the considered threshold that was the accuracy of 88%. As the next point, evaluate function practiced to assess the performance of the trained model. The result of this assessment played a prominent role in ensuring the continuation of the work. This measurement took on the devoted prepared data to this phase. The training procedure ended up utilizing predict function to label ultimate images.

### 3.6 Result

As it is depicted in fig.5, the training scenario will be ended up with classifying our input histopathological microscopy image dataset into three classes. Adopting transfer learning strategy without fine-tuning with taking advantage of pre-trained VGG16 convolutional neural network provide our grading purpose with the validation accuracy of 88%. Considering all of the applied configurations and using Adam optimizer, this significant accuracy obtained after 60 epochs. The related detail to this matter is displayed in fig.6. Some customized DNN models with different number of layers and structure, MobileNet, Inception, and Bag of visual words (BOVW) were other networks which were utilized and tested to get a high accuracy among which VGG16 outperformed all of them. It is also worth mentioning that our introduced model achieved a considerable accuracy of 72% and 2.9 as loss value in the phase of evaluation.

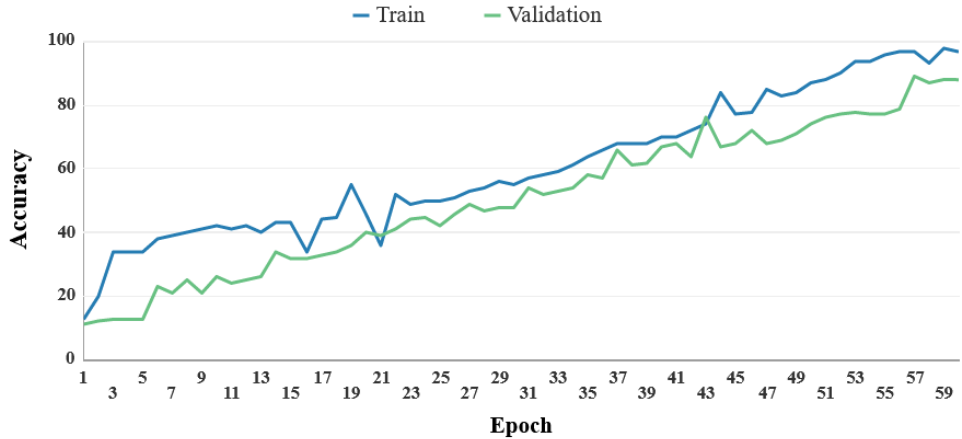

Fig.6 The trend of obtained validation and train accuracy per epoch

## 4. Conclusion

Breast cancer has become the second leading kind of cancer and one of the major causes of death among women. Proper treatment and even prevention from getting this disease stem from early diagnosis. Deep learning plays a prominent role in the detection of this cancer by providing the ability to grade captured images from breast cancer tissues. In this research, we aimed to grade histopathological images from Databiox image dataset whose details are provided in section 3.1. To achieve the desired result, various deep learning structures were examined, namely, some customized DNN models with a different number of layers ranging from 15 to 30, transfer learning technique with diverse pre-trained models with and without fine-tuning. With regard to getting a high accuracy, in the end, we determined a decision to select transfer learning without fine-tuning process as our main strategy in which VGG16 utilized as a feature extractor. The proposed model named BCNet in this study implemented using an integrated API in Keras called tf.keras. Due to the fact that insufficient data size was posed the risk of overfitting on our output and also hamper the performance of our designed model, we adopted the data augmentation method to expand the size of the input image dataset in pre-processing step. Moreover, in order to minimize loss value and enhance the model's training efficiency, an Adam optimizer was applied during the training process. After the accomplishment of the training stage, we could get a considerable validation accuracy of 88%. Since to our knowledge, there are no previous works on the used dataset in this investigation, we were unable to compare our result. But the obtained outcome by our proposed architecture by itself considers noteworthy.

**Abbreviations**

| MI | Microwave Imaging |
|---|---|
| Adam | Adaptive Moment Estimation |
| API | Application Programming Interface |
| APS | Average Precision Score |
| AUC | Area Under the Curve |
| BACH | Breast Cancer Histology Challenge |
| BC | Breast Cancer |
| BGR | Blue, Green and Red |
| BOVW | Bag of visual words |
| BCNet | Breast cancer network |
| CNN | Convolutional Neural Network |
| DNN | Deep Neural Network |
| DPI | Dots Per Inch |
| GAP | Global Average pooling |
| GCN | Global Contrast Normalization |
| H&E | Hematoxylin And Eosin |
| IDC | Invasive Ductal Carcinoma |
| IoT | Interned of things |
| JPEG | Joint Photographic Experts Group |
| LR | Logistic Regression |
| MD | Magnification Dependent |
| MRI | Magnetic Resonance Imaging |
| RCNN | Recurrent Convolutional Neural Network |
| ReLU | Rectified linear Activation Function |
| ResNet | Residual Network |
| RGB | Red Green Blue |
| ROC | Receiver Operating Characteristics |
| SGD | Stochastic Gradient Descent |
| SVM | Support Vector Machine |
| VGG | Visual Geometry Group |
| WTA | Winner Take All |